\documentclass{ifacconf}
\usepackage{stfloats}  % Needed for [!tbp] placement specifier

\usepackage{graphicx}      % include this line if your document contains figures
\usepackage{natbib}        % required for bibliography

\usepackage{amsmath}         %Tillagt av Jonas
\usepackage{mathtools}  % coloneqq
\usepackage{amsfonts}
\usepackage{todonotes}
\usepackage{mathabx}
\usepackage{comment}
\usepackage{ulem}
\usepackage{multirow}
\usepackage{array}
\usepackage{lipsum}
\usepackage{booktabs}

\usepackage{url}
\usepackage{mathtools}
\mathtoolsset{showonlyrefs}

\usepackage{enumitem}

\usepackage{xcolor}

\DeclareMathOperator*{\argmin}{\arg\!\min}

\newcommand{\Rmin}{R_{\min}}           % R_min notation
\newcommand{\Rmax}{R_{\max}}           % R_max notation
\newcommand{\Rcom}{R_{\mathrm{com}}}   % max com. R (when SNR=SNR_0)
\newcommand{\K}{\mathcal{K}}  % agent index set

\newcommand{\tnode}{\mathsf{s}} 
\newcommand{\rnode}{\mathsf{r}}  
\newcommand{\cnode}{\mathsf{c}} 
\newcommand{\jnode}{\mathsf{j}} 

\newcommand{\jammer}{\mathsf{j}}   % jammer index
\newcommand{\B}{\mathbf{B}}  % open ball
\newcommand{\apos}{\delta p}     % action position/velocity/motion
\newcommand{\aori}{\delta \phi}  % action orientation
\newcommand{\maxpos}{\sigma_p} % max velocity
\newcommand{\maxphi}{\sigma_{\phi}} % max antenna angle velocity

\newcommand{\cpos}{c_{\apos}}     % cost position/motion/velocity
\newcommand{\cphi}{c_{\aori}}     % cost antenna orientation 

\newcommand{\jpos}{p_{\jammer}}         % jammer position
\newcommand{\jvel}{\delta p_{\jammer}}  % jammer velocity

\newcommand{\SINR}{\mathsf{SINR}}
\newcommand{\SINRT}{\SINR_0}  % SNR threshold
\newcommand{\pproj}{\bar{p}}  % meetup position of agent (orthogonal projection onto line between retrieval point and receiving base)
\newcommand{\jinit}{\mathbf{C}}  % jammer init capsule

\newcommand{\Herm}{\mathsf{H}} % conjugate transpose

\makeatletter
\providecommand{\thepart}{}

\makeatother

\begin{document}
\begin{frontmatter}

\title{Dynamic one-time delivery of critical data by small and sparse UAV swarms: a model problem for MARL scaling studies\thanksref{footnoteinfo}} 

\thanks[footnoteinfo]{The first author thanks the Wallenberg AI, Autonomous Systems and Software Program (WASP) funded by the Knut and Alice Wallenberg Foundation. Fredrik Båberg, Anders Israelsson and Johan Markdahl at FOI are acknowledged for their great support, Axel Ringh and Ann-Brith Strömberg at Chalmers for careful reading, and reviewer one and two for insightful comments.}

\author[First,Third]{Mika Persson}
\author[Second]{Jonas Lidman}
\author[First]{Jacob Ljungberg} 
\author[First]{Samuel Sandelius}
\author[First,Third]{Adam Andersson} 

\address[First]{Saab AB, 112 76, Gothenburg, Sweden (mikape@chalmers.se)}
\address[Second]{Swedish Defence Research Agency (FOI), 164 90, Stockholm, Sweden (jonas.lidman@foi.se)}
\address[Third]{Chalmers University of Technology and the University of Gothenburg, Department of Mathematical Sciences, 412 58, Gothenburg, Sweden (mikape@chalmers.se)}

\begin{abstract}                

This work studies the application of Multi-Agent Reinforcement Learning (MARL) to decentralized control of unmanned aerial vehicles to relay a critical data package to a known position.
For this purpose, a family of deterministic games is introduced, designed for MARL scaling studies. 
A robust baseline policy is proposed which restricts agent motion and applies Dijkstra's shortest path algorithm. 
Computational experiment results show that two off-the-shelf MARL algorithms perform competitively with the baseline for a small number of agents, but face scalability issues as the number of agents increases. Source code and animations are available online at \url{https://github.com/mikapersson/Information-Relaying}.

\end{abstract}


\end{frontmatter}
%===============================================================================

\section{Introduction}

Consider a search mission in which multiple Unmanned Aerial Vehicles (UAVs) survey a designated area with the goal of locating targets of interest and collecting associated data. The mission involves dynamic UAVs, static base stations, and entities that may interfere.
After the search concludes and a base station or UAV obtains critical data, the scattered UAV swarm initiates a coordinated task to swiftly deliver the data to a base station at a known location. This return phase is the topic of the current work. The UAVs can communicate and control their motion. Moreover, the UAV swarm is sparse, meaning that the UAVs cannot generally form a static connected communication chain. Instead, they must physically move to relay and deliver data, similarly to a rugby team.

In this work, a family of deterministic games is introduced that models the described problem and is suitable for scalability studies in Multi-Agent Reinforcement Learning (MARL). 
The formulation captures key elements of the real problem while introducing simplifications, most notably the assumption of perfect information. A handcrafted and well-performing baseline policy is introduced, and two MARL methods from the literature are trained and evaluated on scenarios involving up to nine UAVs. 
The latter are Multi-Agent Proximal Policy Optimization (MAPPO) \citep{MAPPO} and Multi-Agent Deep Deterministic Policy Gradient (MADDPG) \citep{MADDPG}. 
These methods are known for having good learning properties although MARL is known for having scaling problems in the number of agents, see, e.g., \cite{gronauer2022multi}. 
Four scenarios were evaluated, corresponding to the combination of isotropic or directed data links with the presence or absence of a jammer.

To the best of the authors' knowledge, the problem of delivering one single data package with UAVs has not been previously reported on. 
The use of UAVs to maintain resilient data links in relay networks is, however, well studied \citep{Bai2023}. A vast part of the literature consists of civilian applications such as cellular networks and mobile edge computing. 
\cite{Zhang2020} investigate communication via relay UAVs under presence of eavesdroppers. The UAV policies are trained using MADDPG and an extended variant, Continuous Action Attention MADDPG. The paper demonstrates successful training of one transmitting UAV together with two jammer UAVs. 
Similarly, \cite{Bai2024} utilize relay UAVs to maintain a secure communication while avoiding eavesdropping by an adversarial agent.
The proposed Covert-MAPPO algorithm is successfully applied to a scenario with two relay UAVs. 
A related problem concerning motion control for communication is presented in \cite{Zhu2021}, where a dispersed UAV swarm aims to merge into a single connected cluster, forming a fully connected communication graph. They introduce the Decomposed MADDPG algorithm, which performs efficiently for swarms of up to twelve UAVs. 
In the mentioned works, novel methods were mostly compared against their respective original versions. This is a somewhat unsatisfactory comparison in settings where the original counterparts lack reliability.
This motivates the introduction of the handcrafted nontrivial baseline described in Section \ref{sec:baseline}. Regarding benchmarks, \cite{pan2022mate} introduces one in which an agent group solves a logistics problem while minimizing exposure to an adversarial surveillance team.

Section \ref{sec:toy_problem} presents the model problem, Section \ref{sec:policies} the baseline policy and MARL algorithms, and Section \ref{sec:experiments} the experimental setup and results of the simulated scenarios. Finally, Section \ref{sec:conclusions} contains conclusions and potential future directions.

\textbf{Notation:} Let $\|\cdot\|$ and $\langle\cdot,\cdot\rangle$ denote the Euclidean norm and scalar product, respectively. The \textit{open ball} centered at $x\in\mathbb{R}^2$ with radius $r>0$ is defined as
$\B(x,r) = \{y\in \mathbb{R}^2\colon \|y-x\|<r\}$. The \textit{convex hull} of a set $B$ is the smallest convex set containing $B$, i.e.,
\begin{equation}
    \mathrm{Conv}(B)
    =
    \{
      \lambda b_1+(1-\lambda)b_2\colon
      b_1,b_2\in B,\ \lambda\in[0,1]
    \}.
\end{equation}
The $n$-fold Cartesian product of a set $A$ is denoted by $A^n$. Let $j=\sqrt{-1}$ denote the imaginary unit and for $v\in\mathbb{C}^n$ let $v^{\Herm}$ denote the conjugate transpose of $v$. Finally, the circular $k$-shift of a sequence $\alpha=(\alpha_1,\dots,\alpha_M)$ of size $M$ is given by 
$\mathrm{shift}(\alpha;k)=(\alpha_{k+1},\dots,\alpha_M,\alpha_1,\dots,\alpha_k)$ for $k\in\{0,\dots,M-1\}$.

\section{A model problem}
\label{sec:toy_problem}
In this section, a family of deterministic games is introduced. The games involve agents that control their motion and antenna orientations in a 2D scene. 
The agents receive a common terminal reward---the \textit{budget}---upon successful delivery of the \textit{message}.
Prior to delivery, the agents incur only negative rewards, corresponding to the cost of their actions. 
To enable a fair scaling study, the budget is chosen to accommodate a variable number of agents and geometries while maintaining the budget-to-total cost ratio.
Sections~\ref{sec:communication_model}--\ref{sec:budget} introduce the communication model, the scene geometry and state transition, and the budget and cost structure, respectively. 
Section~\ref{sec:generalizations} discusses natural extensions.

\subsection{Communication model}
\label{sec:communication_model}
All agents are equipped with separate transmitting and receiving antennas, where the receivers always are isotropic, while the transmitters may be directional.
However, the sending base always transmits isotropically.

\begin{figure}[hbt!]
    \centering
    \includegraphics[width=\columnwidth, trim={1cm 2.8cm 2.8cm 3.5cm},  % left, bottom, right, top
    clip]{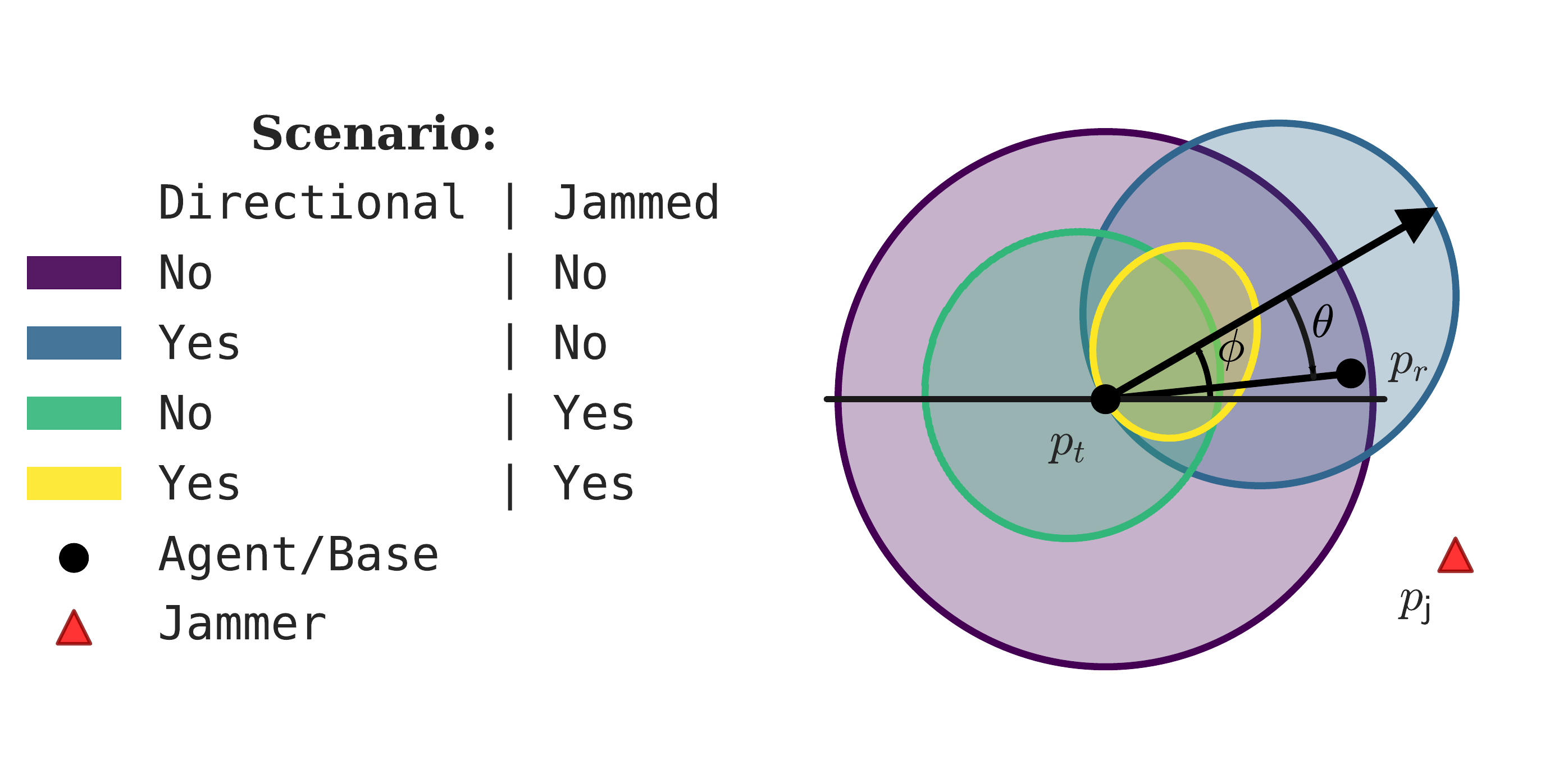}
    \caption{Communication ranges for isotropic and directional transmission with and without jammer. 
    }
    \label{fig:communication_illustration}
\end{figure}
Consider a setting with one transmitter and one receiver, which are either agents or base stations, and one jammer with positions $p_t$, $p_r$, and $p_\jnode$, respectively. Let $\phi\in[0,2\pi)$ denote the antenna orientation of the transmitter, and let $\theta \in[-\pi,\pi)$ be the angle between the transmitter's boresight and the line connecting the transmitter and the receiver (see Fig.~\ref{fig:communication_illustration}). This angle is given by
\begin{equation}
    \theta=\mathrm{mod}(\mathrm{atan2}(p_r-p_t)-\phi +\pi,2\pi)-\pi.
\end{equation}
Based on these quantities, the \textit{Signal-to-Interference-and-Noise-Ratio} (SINR) in the antenna transmission direction, i.e., for $\theta \cdot C_{\mathrm{dir}}\in[-\frac\pi2,\frac\pi2]$, is given by
\begin{equation}\label{eq:snr_general}
  \SINR
  =
  \frac{|\boldsymbol{a}(0)^{\Herm} \boldsymbol{a}(\theta)|}
  {
    \|p_r-p_t\|^2
    (1+C_{\mathrm{jam}}\|p_r-p_\jnode\|^{-2})
  },
\end{equation}
and $\SINR=0$ for $\theta \cdot C_{\mathrm{dir}} \not \in [-\frac\pi2, \frac\pi2]$.
The \textit{steering vector} 
\[
  \boldsymbol{a}(\theta) 
  =
  [
    1, 
    C_{\mathrm{dir}}e^{j\pi \sin(\theta)}
  ]^\top
\]
models directional transmission from a two-element array antenna with half-wavelength spacing when $C_{\mathrm{dir}}=1$, and isotropic transmission when $C_{\mathrm{dir}}=0$. 
For a present jammer, the parameter $C_{\mathrm{jam}}$ is set to $3$, otherwise $C_{\mathrm{jam}}=0$. The presented model is applied to all transmitting agents---including the base stations---throughout our experiments. Communication is successful when $\SINR\geq\SINRT = 1$, where $\SINRT$ is the SINR threshold. This choice results in an isotropic, interference-free communication range of $\Rcom=1$.

\subsection{Fully collaborative deterministic game model}
\label{sec:game}

The game is modeled as an infinite-horizon, fully collaborative deterministic dynamic game with homogeneous agents in discrete time and discounted payoff. Let $\K\!=\!\{1,\hdots,K\}$ denote the agent index set with $K$ agents. A typical \textit{state} at time $t$ is given by
\begin{equation}\label{eq:state}
    x_t
    =
    \big(
    (p_{t,k},\phi_{t,k},b_{t,k})_{k\in\K}
    , 
    p_{t,\mathsf{j}}, \delta p_{t,\mathsf{j}},R,w_t
    \big),
\end{equation}
where $p_{t,k}$ and $\phi_{t,k}$ denote the position and antenna orientation of agent $k$, respectively, and $b_{t,k}$ a Boolean indicating whether the agent is carrying the message. The number $R>0$ is the distance between the sender and receiver base stations located at $p_\tnode=(0,0)$ and $p_\rnode=(R,0)$, respectively. 
%The range $R$ is constant during the game. 
To model the end of the game, a variable $w_t\in\{0,1,2\}$ is used. It takes the value $w_t=0$ before the message is successfully delivered to the receiver base station, $w_t=1$ at the time the receiver obtains the message, and $w_t=2$ thereafter, leading to an absorbing state in which all entities stop and the game terminates. 
The \textit{state space} $X$ is the set of all states $x_t$ of type \eqref{eq:state}. 

At each discrete time step $t$, the agents take actions
\begin{equation}\label{eq:action}
    a_{t,k}=
    (\apos_{t,k},\aori_{t,k}),
    \quad
    k\in \K,
\end{equation}
where $\apos_{t,k}\in\mathbb{R}^2$ is the positional displacement vector and $\aori_{t,k}\in[-\pi,\pi)$ is the orientation displacement. 
The \textit{joint action} at time $t$ is given by $a_t=(a_{t,k})_{k\in\K}$.
Each agent acts within the \textit{action space} $A= \mathbf{B}(0,\maxpos)\times [-\maxphi,\maxphi]$, where $\maxpos$ and $\maxphi$ are the maximum position and antenna angle displacements, respectively. 

The game dynamics are governed by the \textit{transition function} $q\colon X\times A^K\to X$, which maps the current state to a new state given all agent actions.
Given a state $x_t$ of the form~\eqref{eq:state} and agent actions of the form \eqref{eq:action}
for each $k\in \K$, the transition function maps to new agent positions and antenna orientations according to
\begin{equation}\label{eq:state_transition}
  p_{t+1,k}=p_{t,k}+\delta p_{t,k}
  \;\;\;
  \textrm{and}
  \;\;\;
  \phi_{t+1,k}
  =
  \phi_{t,k} + \delta \phi_{t,k}, 
  \;\;\; k\in \K.
\end{equation}
After this update, if an agent $k$ not carrying the message is within communication range of the sending base or any agent $\ell\in\K\setminus\{k\}$ carrying the message at time $t$, then $b_{t,k}=0$ transitions to $b_{t+1,k}=1$, where it remains throughout the game.
The jammer state is updated according to $p_{t+1,\jnode} = p_{t,\jnode} + \delta p_{t,\jnode}$ for $p_{t,\jnode}$ in a convex set $\jinit$ and otherwise $\delta p_{t+1,\jnode} = -\delta p_{t,\jnode}$, making it turn back into $\jinit$. 

\begin{figure}[hbt!]
    \centering
    \includegraphics[
        width=0.94\columnwidth,
        % left, bottom, right, top
        %trim={5pt 0pt 255pt 0pt},
        trim={0.18cm 0cm 8.96cm 0cm},
        clip
    ]{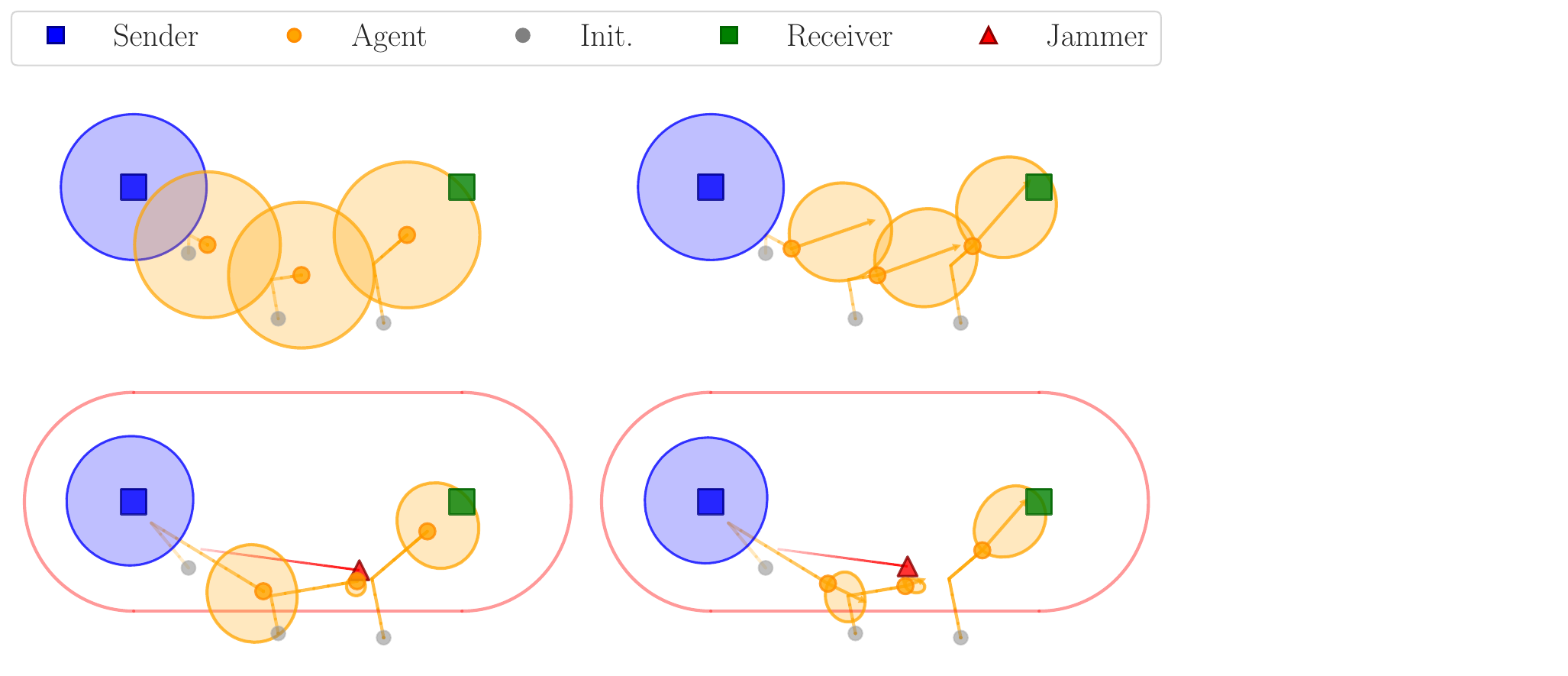}
    \caption{A scene with baseline trajectories and $K=3$ for all scenarios. The red capsule is the jammer area $\jinit$ and the shaded areas represent communication range.}
    \label{fig:baseline_scenario_trajectories}
\end{figure}

The two remaining components of the game are the \textit{one-step reward function} $r\colon X\times A^K\to \mathbb{R}$, which is equal for all agents, and the \textit{discount factor} $\gamma=0.99$. 
Section \ref{sec:budget} describes the reward function in detail.
The game is thus defined by the tuple $(\K, X, A, q, r, \gamma)$.
The objective of the game is to find a stationary policy $\pi_k\colon X\to A$ for each agent $k$, which for every state $x\in X$ maximizes the \textit{value function}
\begin{equation}\label{eq:value_function}
    V_{\pi_k}(x)
    =
    \sum_{t=0}^\infty
    \gamma^t
    r(x_t,a_{t}),
    \quad
    a_{t}=(\pi_k(x_t))_{k\in\K},
\end{equation}
where $x_t\!=\!q(x_{t-1},a_{t-1})$ and $x_0\!=\!x$. 
The time index is suppressed if no ambiguity arises.
An example scene with three agents is illustrated in Fig.~\ref{fig:baseline_scenario_trajectories}.

\subsection{Action budget and costs}
\label{sec:budget}
\begin{figure}[hbt!]
    \centering      % <left> <bottom> <right> <top>
    \includegraphics[width=\columnwidth, trim={0cm 0.35cm 0cm 0cm}, clip]{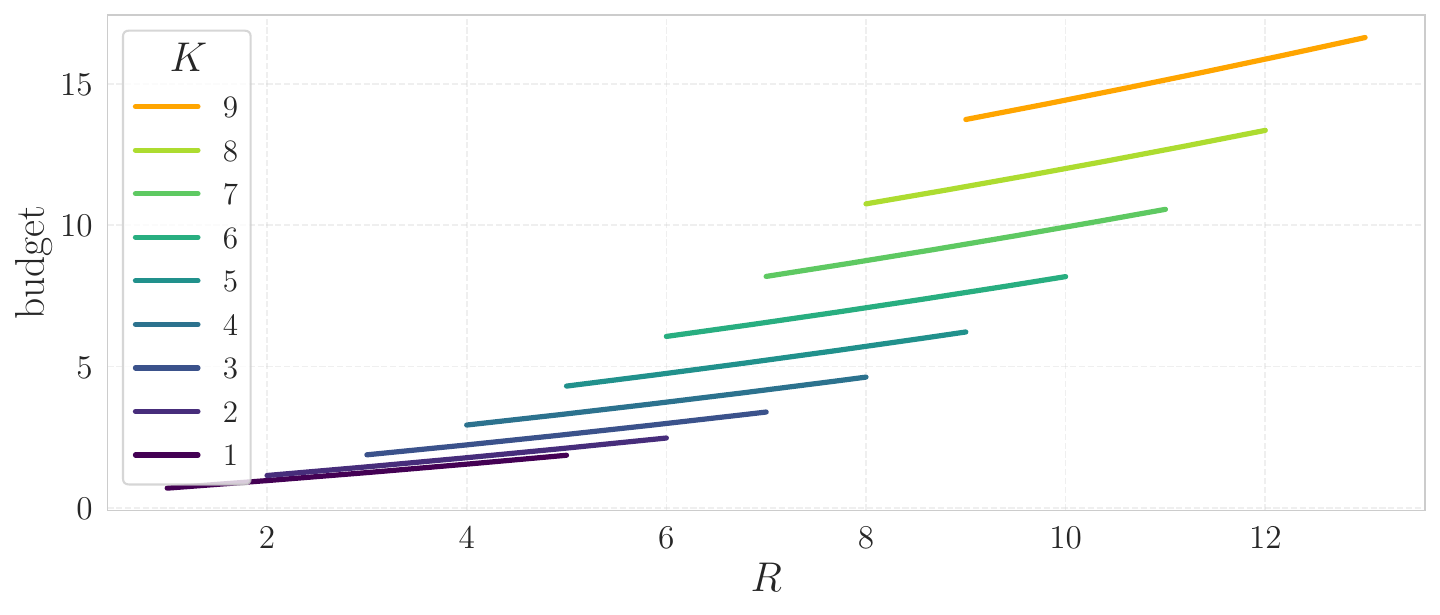}
    \caption{The budget for different $R$ and $K$.}
    \label{fig:budget}
\end{figure}

\begin{figure}[t]
    \centering
    \includegraphics[width=\columnwidth]{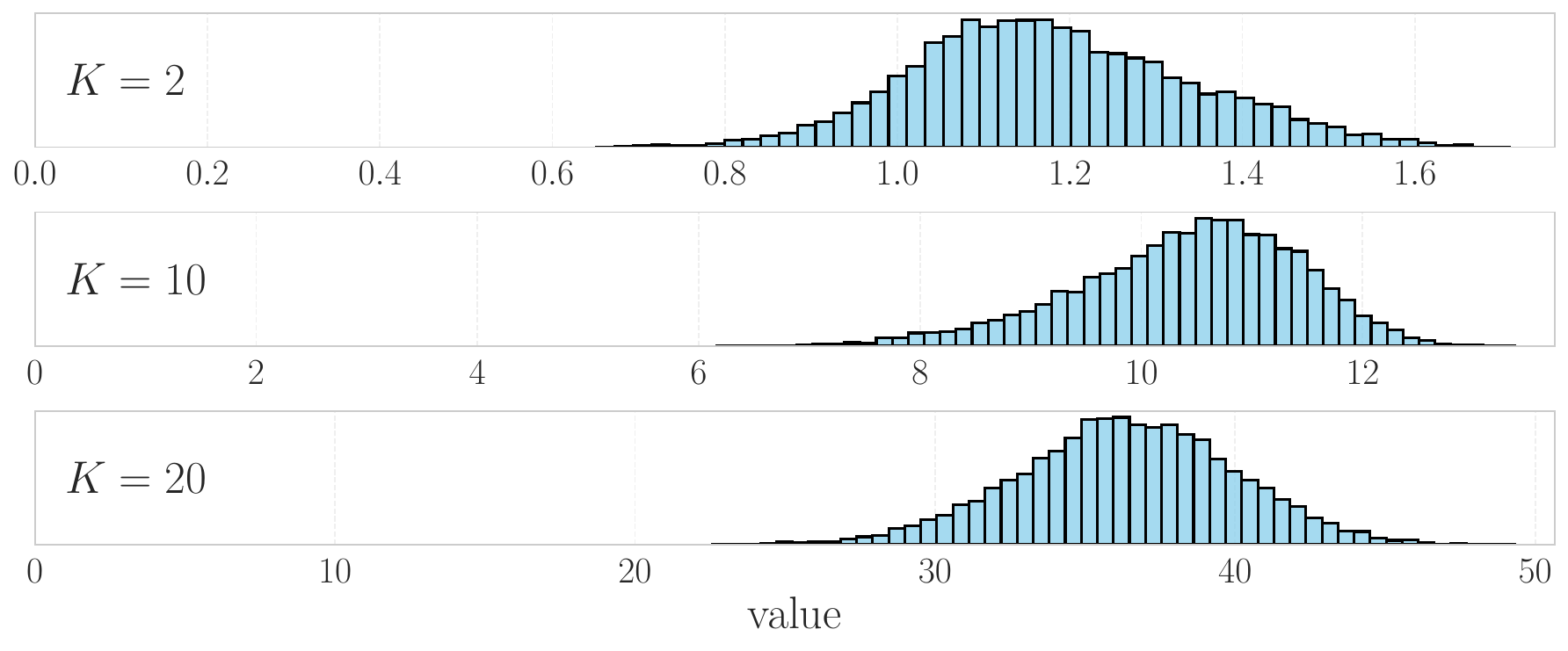}
    \caption{Values for the isotropic non-jammed scenario, the baseline policy of Section~\ref{sec:baseline}, $K=2,10,20$ agents, and 10'000 initial states.}
    \label{fig:value_histogram_compare}
\end{figure}

The shared reward function of all agents is given by
\begin{equation*}\label{eq:reward}
    r(x,a)
    \!
    =
    \!
    \mathrm{budget}(R,w;K)
    \!
    -
    \!
    \cpos\sum_{k=1}^K\|\delta p_k\|^2
    \!
    -
    \!\cphi\sum_{k=1}^K|\delta \phi_k|^2,
\end{equation*}
where $\cpos,\cphi\geq0$ are fixed parameters. The budget is non-negative and is given at the time of delivery of the message, i.e., for $w=1$. 
To obtain a budget with the desired scaling in the number of agents, the dimensioning state $x_\sharp$ is used in which all agents start in position $p_\sharp=(1.1R,0)$, i.e., 
%they are positioned 
behind the receiver base. 
This state requires considerable agent movement, and the budget is set as the discounted accumulated motion cost. 
The budget is computed to solve the game with isotropic transmission and no jammer, starting from $x_\sharp$ using the baseline policy (see Section~\ref{sec:baseline}). 
To clarify, the budget is chosen to ensure that agents have a reasonable opportunity to obtain a positive value --- corresponding to delivering the message --- from any initial state. If the budget is too low, the value function would be maximized by all agents remaining passive.
In detail, $\mathrm{budget}(R,0;K) = \mathrm{budget}(R,2;K) = 0$
and
\begin{equation*}
    \mathrm{budget}(R,1;K)
    =
     \frac{1}{ \gamma^{T_\sharp}}
        \sum_{t=0}^{T_\sharp-1}
        \gamma^t
        \sum_{k=1}^K
        \|\delta p_{k,t}\|^2,
\end{equation*}
where $\delta p_{k,t}$ is the movement of agent $k$ at time step $t$ and $T_\sharp$ is the delivery time. The division by $\gamma^{T_\sharp}$ is a means to reduce the discount of the terminal reward. 
The budget has some small, undesired fluctuations in $R$ due to the discrete nature of the game; 
to smooth it, a second-order polynomial is fitted and used (see Fig.~\ref{fig:budget}). For simplicity, this budget is also used for scenarios with directional transmission and the presence of a jammer.
The parameters $\cpos,\cphi\in[0,1)$ are chosen sufficiently small so that---with a good margin---the value $V_\pi(x_\sharp)$ using $K$ agents with the baseline policy exceeds the corresponding value for any smaller number of agents. 
In this work, $\cpos=0.5$ and $\cphi=0.1$ are chosen. 
Fig.~\ref{fig:value_histogram_compare} shows that the shape of the value distribution remains largely invariant with respect to the number of agents. Furthermore, aside from a scaling factor, the margin above zero is substantial, as desired.

\subsection{Discussion about generality}
\label{sec:generalizations}

Several convenient design choices were made in the proposed model problem. A deterministic game was chosen to allow for a simple and very fast value evaluation, although this is not necessary. Adding noise to the state transition in position and orientation is both straightforward and natural, turning the game into a Markov game. Noise often has a regularizing effect in optimal control, and it might improve the training of MARL. Limited tests with noise were made, but no significant difference was observed.

Another choice was to employ a reward in which every agent is penalized for the actions of all other agents. This choice was made to make every agent care for the energy consumption of the entire swarm. An alternative would be to let each agent receive costs only for its own actions; this was not investigated. Moreover, while homogeneous agents allow a joint policy network, having heterogeneous agents is also possible, but requires multiple policy networks. Regarding directional transmission, it is straightforward to increase directivity by having $L\geq2$ antenna elements that lead to the steering vector
\[
  \boldsymbol{a}(\theta) 
  =
  [
    1, 
    e^{j2\pi \sin(\theta)/L},
    ...,
    e^{j2\pi (L-1)\sin(\theta)/L}
  ]^\top. 
\]
Finally, the jammer dynamics can be extended to something more challenging, either hard-coded to approach agents, or as an adversarial learning agent of the game.

\section{Policies under investigation}
\label{sec:policies}

The problem of delivering a message as quickly as possible while ignoring costs is a complex optimization problem involving three intertwined key decision problems: 
\begin{enumerate}[label=\Roman*)]

    \item \label{factor-I} selecting which agents should participate in the relay,

    \item \label{factor-II} finding the order in which these agents should relay the message, and
    
    \item \label{factor-III} determining the relay points at which message transfers occur.
\end{enumerate}

Each agent selected in \ref{factor-I} has a \textit{retrieval point} and a \textit{handover point} at which it receives and transfers the message, respectively. These are determined in \ref{factor-III} and collectively referred to as \textit{relay points}.
The corresponding \textit{relay times} are the time instants at which inter-agent transfers occur. The \textit{motion envelope} of an agent is the set of positions it can reach in the time it can take to satisfy its relay constraint. 
Let $k_i\in \K$ be the $i^{\text{th}}$ agent to receive the message. 
If the preceding agents $k_1, \ldots, k_{i-1}$ act suboptimally, their delayed relay times enlarge the motion envelope of agent $k_i$,
while near-optimal decisions by preceding agents reduce these relay times, leading to a smaller motion envelope for agent $k_i$. 
This interdependence between the decisions of preceding agents, together with the combinatorial complexity of the problems in \ref{factor-I}--\ref{factor-II}, highlights the inherent difficulty of solving the problem efficiently. 
The baseline computes centralized open-loop policies from the initial state and executes them without feedback, whereas the MARL approach uses centrally trained, decentralized feedback policies.
Sections~\ref{sec:baseline}--\ref{sec:marl} presents a handcrafted baseline and two standard MARL methods, respectively.

\subsection{A baseline controller}
\label{sec:baseline}
The following description assumes an isotropic, non-jammed scenario.
The baseline controller reduces the dynamic optimization problem into a static graph problem by first determining the agent relay order in \ref{factor-II} and then restricting the agents' motion envelopes in \ref{factor-III}. 
This lends a static graph formulation in which the problems in \ref{factor-I} and \ref{factor-III} can be solved using Dijkstra's algorithm, which is executed for each of the $K$ agents as the (initial) retrieving agent, and the best of the $K$ solutions is chosen. 
Agents included in the best path form a \textit{relay chain}, thereby solving \ref{factor-I} and \ref{factor-II}, while the remaining agents are \textit{passive agents}. 

\begin{figure}[t]
    \centering
    \usetikzlibrary{calc,intersections,through,backgrounds, math}

%\begin{document}

\begin{tikzpicture}[scale=1]

% Parameters
%\def\Rcom{1}                  % circle radius
\pgfmathsetmacro{\Rcom}{1}
\def\R{6}                  

% Draw bases
\node[circle, fill=black, inner sep=1.5pt, label={[label distance=0.0mm]left:$p_{\tnode}$}] (TX) at (0,0) {};
\draw (TX) circle (\Rcom);

\node[circle, fill=black, inner sep=1.5pt, label={[label distance=0.0mm]right:$p_{\rnode}$}] (RX) at (\R,0) {};
\draw (RX) circle (\Rcom);

% Draw retrieving agent
\node[circle, fill=black, inner sep=1.5pt, label={[label distance=0.0mm]below:$p_k$}] (retriever) at (0.5\Rcom,-2.5\Rcom) {};
%\draw (retriever) circle (\Rcom);

% Relay agent
\node[circle, fill=black, inner sep=1.5pt, label={[label distance=0.0mm]left:$p_i$}] (relay_init) at (3.35\Rcom,-3.3\Rcom) {};
%\draw (relay_init) circle (\Rcom);

% Draw line to retrieving point
% Compute distance and store in macro
% \tikzmath{coordinate \temp;
% %Storing coordinates difference
% \temp = (TX)-(retriever); 
% %Computing the length of C = (Cx,Cy) from its components Cx and Cy
% %Note the length \distAB is in points (pt)
% \distAB = sqrt((\tempx)^2+(\tempy)^2); 
% \rat = (\distAB - \Rcom)/\distAB
% }

% \path let
%   \p1 = (retriever),
%   \p2 = (TX)
% in
%   \pgfextra{
%     \pgfmathsetmacro{\distAB}{veclen(\x2 - \x1, \y2 - \y1)}
%     \pgfmathsetmacro{\rat}{(\distAB - \Rcom)/\distAB}
%     \edef\rat{\rat} % fully expand here
%   };

% RETRIEVAL POINT
% \def\rat{0.615} 
% \node[
%   circle, 
%   fill=black, 
%   inner sep=1.5pt, 
%   label={[label distance=0mm]above left:{$p_{\tnode_k}$}} 
% ] 
% (retrieve_point) at ($ (retriever)!\rat!(TX) $) {};
\node[circle, fill=black, inner sep=1.5pt, label={[label distance=-1.5mm]above left:$\hat{p}_{k}$}] (retrieve_point) at (0.5\Rcom,-0.85\Rcom) {};

% Draw line from retrieving agent to retrieving point
%\draw (retriever) -- (retrieve_point);
\draw[|-|] 
  (retriever) 
  -- 
  node[fill=white, inner sep=0.5mm] {\small $c$} 
  (retrieve_point);
%\coordinate [label=right:]  () at ($(retriever)!.5!(retrieve_point) $) {};

% Draw line from retrieving point to receiving base
\draw[dashed] (retrieve_point) -- (RX);

% Draw line from relay agent to its projection point
\def\rat{0.422}
\node[
  circle, 
  fill=black, 
  inner sep=1.5pt, 
  label={[label distance=0mm]above right:{$\bar{p}_i$}} 
] 
(proj_relay) at ($ (retrieve_point)!\rat!(RX) $) {};
\draw[dashed] (relay_init) -- (proj_relay);

% Mark relay point
\node[circle, fill=black, inner sep=1.5pt, label={[label distance=0.0mm]left:$q_i$}] (relay) at ($ (relay_init)!.65!(proj_relay) $) {};
\draw (relay) circle (\Rcom);
\draw[->] (relay_init) -- (relay);

% Draw line from retrieval point to relay point
\def\rat{0.62}
\coordinate (ret_to_rel) at ($(retrieve_point)!\rat!(relay) $) {};
%\draw[->] (retrieve_point) -- (ret_to_rel);

% Draw unit arrows
% u from retrieval point to RX
\def\rat{0.17}
\coordinate (u) at ($(retrieve_point)!\rat!(RX) $) {};
\draw[->] (retrieve_point) -- (u);
\node[below right] at (u) {$u$};

% v from projection point to relay agent
\def\rat{0.55}
\coordinate (v) at ($(proj_relay)!\rat!(relay) $) {};
\draw[->] (proj_relay) -- (v);
\node[left] at (v) {$v$};

% Draw help lines (a, b, d, lambda, and delta constants)
% a
\def\perpoffset{-3mm}
\path let
  \p1 = (relay_init),
  \p2 = (proj_relay),
  \n1 = {veclen(\x2-\x1,\y2-\y1)}
in
  coordinate (shifted_A) at ($ (relay_init) + ( {\perpoffset*(\y1 - \y2)/\n1}, {\perpoffset*(\x2 - \x1)/\n1} ) $)
  coordinate (shifted_B) at ($ (proj_relay) + ( {\perpoffset*(\y1 - \y2)/\n1}, {\perpoffset*(\x2 - \x1)/\n1} ) $);

\draw[|-|] 
  (shifted_A) 
  -- 
  node[fill=white, inner sep=0.5mm] {\small $a$} 
  (shifted_B);

% c
%\coordinate [label=above:$d$] (d) at ($(retrieve_point)!.5!(proj_relay) $) {};
% \def\perpoffset{3mm}
% \path let
%   \p1 = (retriever),
%   \p2 = (retrieve_point),
%   \n1 = {veclen(\x2-\x1,\y2-\y1)}
% in
%   coordinate (shifted_A) at ($ (retriever) + ( {\perpoffset*(\y1 - \y2)/\n1}, {\perpoffset*(\x2 - \x1)/\n1} ) $)
%   coordinate (shifted_B) at ($ (retrieve_point) + ( {\perpoffset*(\y1 - \y2)/\n1}, {\perpoffset*(\x2 - \x1)/\n1} ) $);
  
% \draw[|-|] 
%   (shifted_A) 
%   -- 
%   node[fill=white, inner sep=0.5mm] {\small $c$} 
%   (shifted_B);

% d
%\coordinate [label=above:$d$] (d) at ($(retrieve_point)!.5!(proj_relay) $) {};
\def\perpoffset{3mm}
\path let
  \p1 = (retrieve_point),
  \p2 = (proj_relay),
  \n1 = {veclen(\x2-\x1,\y2-\y1)}
in
  coordinate (shifted_A) at ($ (retrieve_point) + ( {\perpoffset*(\y1 - \y2)/\n1}, {\perpoffset*(\x2 - \x1)/\n1} ) $)
  coordinate (shifted_B) at ($ (proj_relay) + ( {\perpoffset*(\y1 - \y2)/\n1}, {\perpoffset*(\x2 - \x1)/\n1} ) $);
  
\draw[|-|] 
  (shifted_A) 
  -- 
  node[fill=white, inner sep=0.5mm] {\small $d$} 
  (shifted_B);
  
% lambda
\def\perpoffset{-12mm}
\path let
  \p1 = (relay),
  \p2 = (proj_relay),
  \n1 = {veclen(\x2-\x1,\y2-\y1)}
in
  coordinate (shifted_A) at ($ (relay) + ( {\perpoffset*(\y1 - \y2)/\n1}, {\perpoffset*(\x2 - \x1)/\n1} ) $)
  coordinate (shifted_B) at ($ (proj_relay) + ( {\perpoffset*(\y1 - \y2)/\n1}, {\perpoffset*(\x2 - \x1)/\n1} ) $);
  
\draw[|-|] 
  (shifted_A) 
  -- 
  node[fill=white, inner sep=0.5mm] {\small $\lambda$} 
  (shifted_B);

% delta
\draw[|-|] 
  (retrieve_point) 
  -- 
  node[fill=white, inner sep=0.5mm] {\small $b$} 
  (ret_to_rel);

% Rcom
\draw[|-|] 
  (RX) 
  -- 
  node[fill=white, inner sep=0.5mm] {\small $R_{\mathrm{com}}$} 
  (\R, -\Rcom);

% R
\def\perpoffset{3mm}
\path let
  \p1 = (TX),
  \p2 = (RX),
  \n1 = {veclen(\x2-\x1,\y2-\y1)}
in
  coordinate (shifted_A) at ($ (TX) + ( {\perpoffset*(\y1 - \y2)/\n1}, {\perpoffset*(\x2 - \x1)/\n1} ) $)
  coordinate (shifted_B) at ($ (RX) + ( {\perpoffset*(\y1 - \y2)/\n1}, {\perpoffset*(\x2 - \x1)/\n1} ) $);
  
\draw[|-|] 
  (shifted_A) 
  -- 
  node[fill=white, inner sep=0.5mm] {\small $R$} 
  (shifted_B);

% Draw both "solutions" of lambda

%\draw (A) -- (B);

%\node [draw,circle through=(B)] at (A) {};

%\draw (A) circle (\Rcom);

\end{tikzpicture}
%\end{document}
    \caption{Illustration of the geometry behind the baseline graph weights. The dashed line represents $L_k$.}
    \label{fig:dijkstra_weight_derivation}
\end{figure}
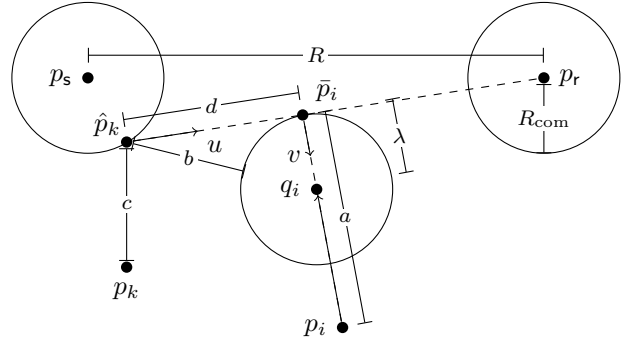

Let $k\!\in\!\K$ be the retrieving agent, and the initial positions of the agents be $p_1,\dots,p_K$.
The \textit{retrieval point} $\hat{p}_{k}$ of agent $k$ is chosen to be the optimal retrieval point for the corresponding one-agent game. 
More precisely, 
\[\hat{p}_{k}=\argmin_{p\in \B(p_\tnode, \Rcom)} \{ \|p - p_k\|+\|p_\rnode - p \| \}\]
if $p_k$ is outside the communication range of $p_\tnode$, otherwise $\hat{p}_{k}=p_k$. 
Fig.~\ref{fig:dijkstra_weight_derivation} illustrates the geometry.
The line segment between the retrieval point and the receiver base station is denoted $L_k$.
In a first step, the movements of the agents are restricted to a perpendicular motion at maximum speed towards $L_k$.
The rationale behind this is that agents on $L_k$ form a straight chain from the retrieval point to the receiving base station. 
Some agents may not reach $L_k$ before they become obsolete, while some agents are close enough to $L_k$ or to other agents so that they still become part of the resulting solution. 
The latter also requires knowledge of relay times, which in turn depend on the message's previous path. 
To avoid such dependencies, conservative bounds on relay times are used. 

Since agents are restricted to perpendicular movement relative to $L_k$, the relaying order is fixed. 
Given this order, the smallest motion envelope of agent $i\in \K_{-k}$, where $\K_{-k}=\K \setminus \{ k \}$, is obtained if all preceding agents are able to distribute themselves uniformly on the line between $\hat{p}_{k}$ and the candidate retrieval point $\hat{p}_i$ for agent $i$. 
To derive $\hat{p}_i$ using this principle, let 
\[\pproj_i=\hat{p}_{k}+\langle p_i - \hat{p}_{k}, u\rangle u\] 
be the projected position of agent $i$ onto $L_k$, where $u = (p_\rnode-\hat{p}_{k})/ \| p_\rnode-\hat{p}_{k} \|$.
For presentation purposes, the agent indices are reordered so that $k=1$ and $\hat{p}_1,\dots,\hat{p}_K$ correspond to the potential retrieval points. 
The sum of distances traveled by the retrieving agent $k$ and the \hbox{$i-2$} subsequent agents carrying and relaying the message to agent $i$ at $\bar{p}_i$, is at least 
\[
    D
    =
    \|\hat{p}_{k}-p_k\| 
    + 
    \max(0,\|\pproj_i-\hat{p}_{k}\| - i\Rcom).
\]
%where $\Rcom$ is the communication distance.
If $\|\pproj_i-p_i\|\leq D$, then agent $i$ can reach $\pproj_i$ in time to receive the message without delay, regardless of the actions of the agents preceding $i$. 
In this case $\hat{p}_i=\bar{p}_i$, and otherwise $\hat{p}_i=q_i$, where $q_i$ is the point on the straight line segment between $p_i$ and $\bar{p}_i$ that solves the equation
\begin{equation}\label{eq:relay_condition}
    \|q_i-p_i\| = \|\hat{p}_{k}-p_k\| + \max(0,\|q_i-\hat{p}_{k}\| - i\Rcom).
\end{equation}
Parameterizing the solution of \eqref{eq:relay_condition} by  $q_i=\pproj_i+\lambda v$, where $v=(p_i-\pproj_i)/\|p_i-\pproj_i\|$ is the unit normal from $L_{k}$ to $p_i$, and $\lambda\geq0$ yields $b = \max \big(0, \sqrt{d^2 + \lambda^2} - i\Rcom\big)$, where
 \begin{align*}
    \lambda 
    &=
    \begin{cases}
        a - c,  & 
        \text{if } 
             \sqrt{d^2\! +\! (a \!-\! c)^2} \!\geq \!i\Rcom,
        \\
        \frac{(a - c + i \Rcom)^2 - d^2}{2(a - c + i \Rcom)}, 
        &
        \text{otherwise};
    \end{cases}
    \\
     a &= \|\pproj_i - p_i\|
     ; \quad
     c = \|\hat{p}_{k}-p_k\|;
     \quad
     d = \|\pproj_i - \hat{p}_{k}\|.
 \end{align*}  

Once the sub-optimal candidate retrieval points $\hat{p}_{1},\hdots, \hat{p}_{K}$ are determined, a weighted graph is constructed. 
Dijkstra's algorithm is subsequently employed to determine the optimal path through the graph. 
The graph is defined as follows:
Fix $k \in \K$ and let $G_k(V,E_k)$ be the graph with node set $V=\{\tnode, \rnode\}\cup \K$, with $\tnode$ and $\rnode$ representing the sending and receiving base station, respectively. 
The edge set $E_k$ is defined as
\begin{align*}
    E_k 
    =
    \{ (k,\tnode), (\tnode,\rnode) \} 
    &
    \cup 
    \!
    \displaystyle 
    \bigcup_{i \in \K_{-k}} 
    \!
    \bigg(
    \!
    \{ (\tnode,i), (i,\rnode) \}
    \cup
    \!
    \bigcup_{\ell \in \{ \K_{-k} \}_{-i}}
    \!
    \{ (i, \ell) \}
    \!
    \bigg)
    ,
\end{align*}
with edge weights representing distances between the corresponding base stations and agents. 
The five types of edges have weights given by 
\begin{align*}
w_k(k,\tnode) 
&= \|\hat{p}_{k}-p_k\|, &&
\\
    w_k(\tnode, \rnode) 
    & = 
    \|p_\rnode - \hat{p}_{k}\|, 
    \\
    w_k(\tnode, i) 
    & = 
    \max(0, \|\hat{p}_i-\hat{p}_{k}\|-\Rcom) ,
    && 
    i\in \K_{-k}
    , 
    \\
    w_k(i, \rnode) 
    & = 
    \max\big(0, \|p_\rnode - \hat{p}_i\| - R_{\text{com}}\big), 
    && 
    i \in \K_{-k},
    \\
    w_k(i, \ell) 
    & = 
    \max\big(0, \|\hat{p}_\ell - \hat{p}_i\| - R_{\text{com}}\big), 
    && 
    i, \ell \in \K_{-k},\ i \neq \ell
    .
\end{align*}

Running Dijkstra's algorithm from agent $k$ to the receiving base $\rnode$ on the graph $G_k$ results in a shortest delivery path $\nu_k$ with the total message carrying distance
\[
    D_{\nu_k} = \sum_{(i,\ell)\in \nu_k} w_k(i,\ell).
\]
The optimal path, $\nu_{k_*}$, is the path with the retrieving agent $k_*$ giving smallest $D_{\nu_{k_*}}$. 
A second run of Dijkstra's algorithm is executed after removing the passive agents from the node set. This may yield an improved solution due to expanded agents motion envelopes

A repulsion mechanism is used that allows agents to spread along $L_k$ and approach handover points at non-perpendicular angles. Its impact is significant when relay points end up in dense clusters so that agents cannot utilize their communication range well. First, the agents are partitioned into clusters, consisting of agents having a maximum distance $\Rcom$ to at least one other agent in the cluster. 
Second, the \textit{remaining movement budget} of agent $i\in\K_{-k}$ is given by 
\[
    d_i  =  \|\hat{p}_{k}-p_k\|+\max(0,\|\hat{p}_i-\hat{p}_{k}\|-i\Rcom) - \|\hat{p}_i-p_i\|.
    \]
An agent is a \textit{fixed agent} if its remaining budget is zero, and a \textit{movable agent} otherwise. 
Each movable agent adjusts its relay point until it runs out of budget, is at distance $\Rcom$ from the closest member of the cluster, or is positioned between two fixed agents. 
Clusters may merge during the process, requiring an iterated repulsion procedure until no new clusters form. Finally, Dijkstra's algorithm is executed once more to finalize the relay path.

For jammed scenarios, the retrieving agent moves directly towards the sending base station, and agents continue their movements beyond their handover points until the message is received. For directional transmission, agents defer antenna steering toward its handover position in order to reduce its discounted steering cost.

\subsection{Multi agent reinforcement learning}
\label{sec:marl}

In addition to the baseline, two MARL algorithms have been utilized for training policies on the various scenarios: Multi-Agent Proximal Policy Optimization (MAPPO) \citep{MAPPO}, and Multi-Agent Deep Deterministic Policy Gradient (MADDPG) \citep{MADDPG}.
Both algorithms are designed to address a central challenge in MARL: the non-stationarity of the environment induced by multiple simultaneously learning agents.
They do this by employing a centralized training with decentralized execution framework, where each agent learns a policy (actor) while relying on a centralized critic that conditions on the joint observation and actions of all agents. 
By providing the critic with this global information, the learning problem becomes stationary, stabilizing training despite continually changing agent policies. 
Given that the environment is already restricted to perfect information, this assumption is reasonable.
During execution, however, each agent acts solely based on its local observations.

\subsection{Implementation details}
\label{sec:implementation}

BenchMARL \citep{BenchMARL} contains implementations of  MAPPO and MADDPG and was used to train and evaluate the policies. The simulation environment was implemented as described in Section \ref{sec:toy_problem} based on the Multi Particle Environments used in \citep{MADDPG}. During training, an episode ends either when the message is delivered to the receiving base station or after a maximum number of time steps $T_{\max}$, according to
\begin{equation}\label{eq:max_time}
    T_{\max}
    = 
    \big\lceil
      C_\mathrm{time}((1.1\Rmax+2\Rcom)\maxpos^{-1}
      +
      K)
      \big\rceil,
\end{equation}
where $C_\mathrm{time}=1$ during training and $C_\mathrm{time}=1.5$, during evaluation, the latter to allow agents to solve the game with a higher success rate. 

Both MAPPO and MADDPG were trained using reward iteration and curriculum learning \cite{narvekar2020curriculum} in order to mitigate sparse rewards and the credit assignment problem. 
In the reward iteration scheme, agents were initially rewarded for transmitting or receiving a message to or from another agent during the first third of the training episodes. The curriculum learning procedure involved initializing agents close to the line connecting the bases and then gradually expanding the initialization region as training progressed.  

For MAPPO, which generates discrete actions, the motion actions $\delta p_k$ are chosen from the set
\begin{equation}
  \big\{\big(\maxpos\cos(2\pi\ell/8),\maxpos\sin(2\pi\ell/8)\big):\ell = 0,\dots,7\big\}\cup\{\mathbf{0}\},
\end{equation}
where $\maxpos=0.2$ denotes the fixed displacement for motion actions. In scenarios with directed communication, the antenna orientation actions $\aori_k$ are chosen from the set $\{-\sigma_\phi, 0 ,\sigma_\phi\}$, where $\sigma_\phi=\pi/8$. The observation for each agent is based on the relative positions to all other agents, as well as their antenna orientations, base stations, and the jammer. Formally, the observation $o_k$ for MAPPO agent $k\in \K$ is given by
\begin{align*}
    o_k = \big[
    & \Delta p_{\tnode,k}, \Delta p_{\rnode,k} , \Delta\jpos{}_{,k}, \jvel, \phi_k, b_k,  (\Delta p_{i,k}, \phi_i, b_i)_{i\in \K_{-k}}
    \big]^\top,
\end{align*}
where $\Delta p_{i,k}=p_i-p_k$ denotes the relative position with respect to other agents or base stations ($i \in \{ \tnode, \rnode \}\cup \K_{-k}$), and the agents $i\in \K_{-k}$ are ordered by ascending distance from agent $k$.
For MADDPG, the observation of agent $k \in \K$ is given by
\begin{align*}
    o_k = \big[&\mathrm{shift}((\Delta p_{1,k}, \dots,\Delta p_{K,k},\Delta p_{\tnode,k},\Delta p_{\rnode,k})_{- k});k-1),\\
    &\qquad \qquad \mathrm{shift}((b_1,b_2,\dots,b_k);k-1)\big]^\top.
\end{align*}
This observation is somewhat unusual because base stations are not explicitly distinguished from agents. The trained network likely infers which positions correspond to the base stations. Surprisingly, this observation, which was found on accident, yielded the best performance for MADDPG by a significant margin, outperforming the observation representation used for MAPPO.

\section{Experiments}
\label{sec:experiments}

\subsection{Scene geometry and initialization}
\label{sec:geometry}

This section presents the scene geometry and the distributions used to initialize the game.
The game scene varies with the number of agents $K\in\mathbb{N}$, to facilitate scaling studies of solution methods, and the distance $R$ between base stations.  
The sender and receiver bases are separated by a distance $R\in[\Rmin,\Rmax]$, where $\Rmin=K\Rcom$ and $\Rmax=(K+4)\Rcom$.
For $R\in[K\Rcom,(K+1)\Rcom]$, the agents can form a static connected chain (a dense game). 
For $R\in((K+1)\Rcom,(K+4)\Rcom]$, at least one agent must physically transport the message (a sparse game). 
With the base distance $R$ drawn uniformly from $[\Rmin,\Rmax]$, 25\% of the games are dense and 75\% sparse.

The sender and receiver base stations are located at $p_\tnode=(0,0)$ and $p_\rnode=(R,0)$, respectively, with $p_\cnode$ denoting their midpoint. The $K$ initial agent positions $p_1,\dots,p_K$ are uniformly sampled on the ball $\mathbf{B}(p_\cnode,0.6R)$, with orientations $\phi_1,\dots,\phi_K$ drawn uniformly from $[0,2\pi)$. 
The jammer position $p_\jnode$ is sampled uniformly from a capsule around and between the bases, represented by the convex hull $\jinit=\mathrm{Conv}(\mathbf{B}(p_\tnode,1.5\Rcom)\cup\mathbf{B}(p_\rnode,1.5\Rcom))$. The jammer displacement vector $\delta p_\jnode$ is of length $\sigma_\jnode=0.1$, with an angle drawn uniformly from the semicircle centered at $p_\jnode$ and directed toward $p_\cnode$. 

\subsection{Results}
In the comparison of the three methods, three performance figures are used: 
the value $V$, the delivery time $T_{\mathrm{del}}$, and the total distance traveled by the agents $D_{\mathrm{tot}}$. 
Because the system is deterministic, success or failure can be determined at any time. If a state leads to failure, the baseline, which guarantees delivery on time, can be used. 
Experiments are conducted for $K=1,3,5,7,9$ agents. As a qualitative complement to quantitative comparison, rollout trajectories for the four scenarios are also displayed for various initial states.

\begin{figure}[b]
    \centering  % <left> <bottom> <right> <top>
    \includegraphics[width=1.0\columnwidth,trim={1.37cm 1.27cm 1.72cm 1.23cm},clip]{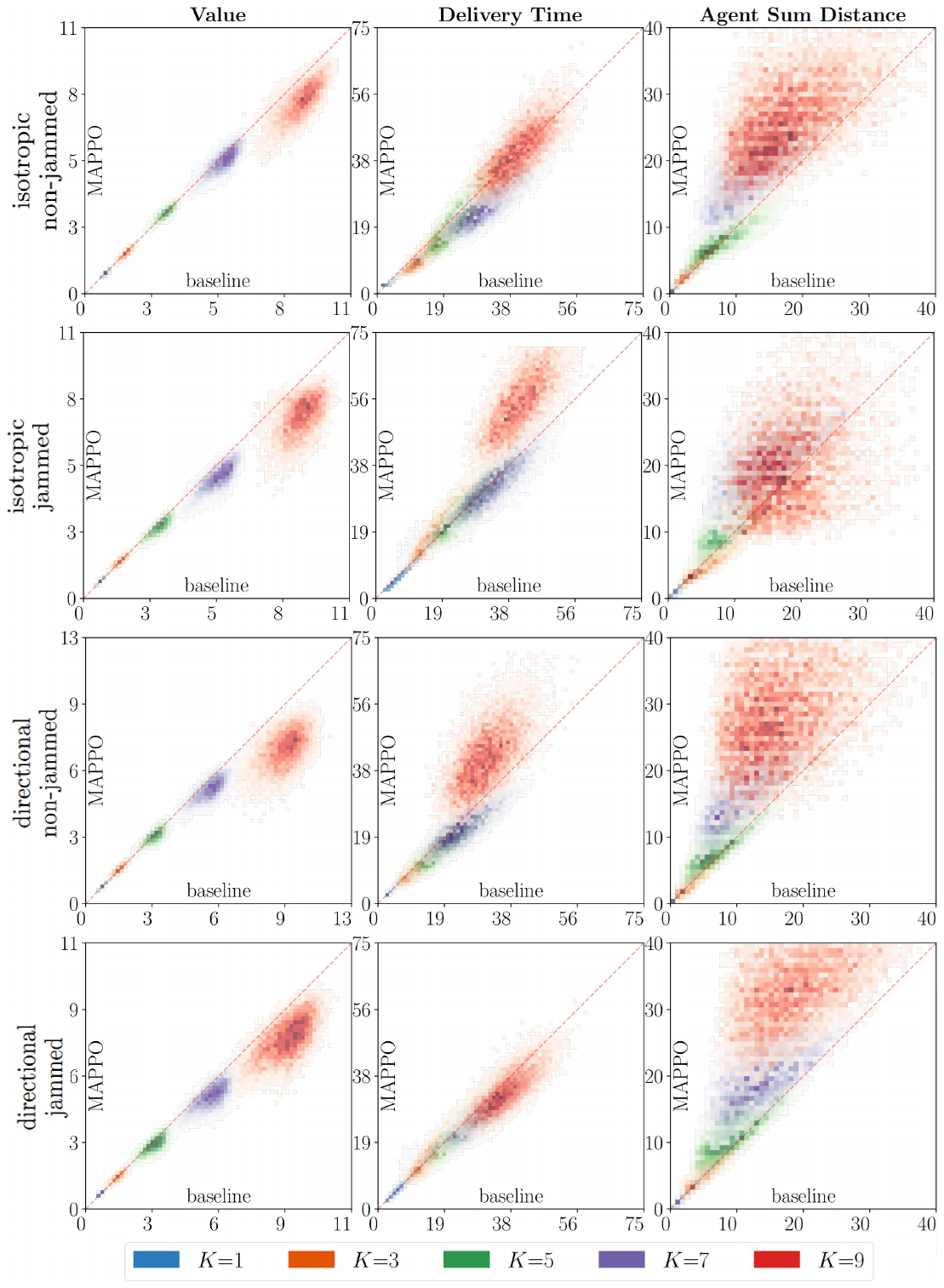}
    \caption{Performance of MAPPO versus baseline, for the four scenarios, $K=1,3,5,7,9$ agents, and 10'000 initial states, best viewed in color.}
    \label{fig:policy_result_comparisons}
\end{figure}

\begin{figure}[t]
    \centering
    \includegraphics[width=\columnwidth]{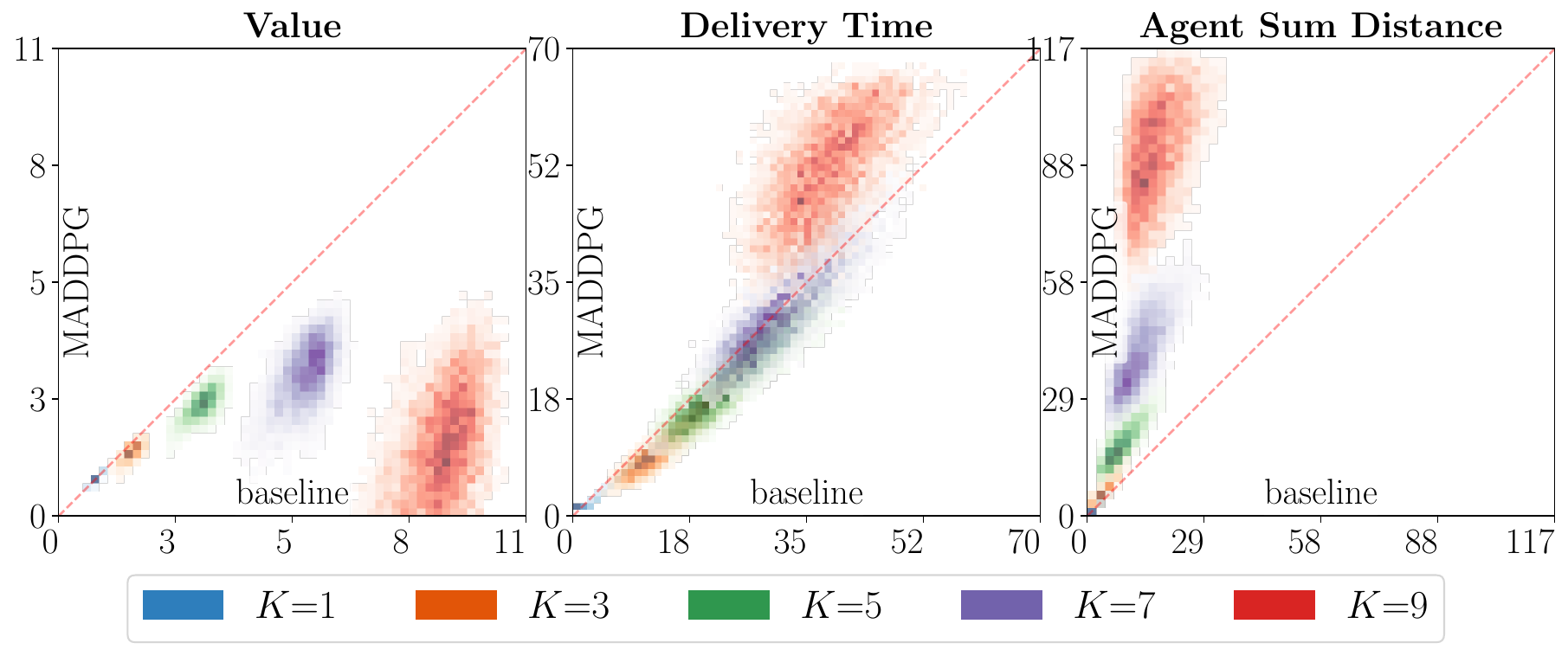}
    \caption{Performance of MADDPG versus baseline, for the isotropic non-jammed scenario, $K=1,3,5,7,9$ agents, and 10'000 initial states, best viewed in color.}
    \label{fig:maddpg_result_comparisons_isotropic_nonjammed}
\end{figure}

First, Fig.~\ref{fig:policy_result_comparisons} presents the pairwise comparisons of the MAPPO and baseline policies for all four scenarios over 10'000 initial states. There is a negligible number of outlier data points with negative values, but these are not shown for presentation reasons. The results show that MAPPO has roughly the same value $V$ as the baseline for $K=1,3,5$, with a small drop for $K=7$ agents and a larger drop for $K=9$. 
For all scenarios, MAPPO generally yields shorter delivery time for $K=1,3,5,7$ agents. For $K=9$ agents, this advantage persists only in the directed-transmission scenario with jamming, while performance is notably worse in the other scenarios.
This discrepancy indicates the randomness of training runs and that repeated trainings have potential to improve the results. 
Fig.~\ref{fig:maddpg_result_comparisons_isotropic_nonjammed} shows that MADDPG achieves values $V$ comparable to the baseline for $K=1,3$ agents. However, the scaling problem become apparent for $K=5$ and severely degrades for $K=7$ and $K=9$. 
Delivery times remain competitive for $K=5$, match the baseline for $K=7$, and again significantly degrade for $K=9$, most probably accompanied by a large cost for unnecessary movement. Due to this poor performance, MADDPG was not evaluated on scenarios other than the isotropic non-jammed case.

\begin{figure}[t!]
    \centering  % <left> <bottom> <right> <top>
    \includegraphics[width=\columnwidth,
    trim={0cm 0.2cm 0cm 0.25cm}, clip]{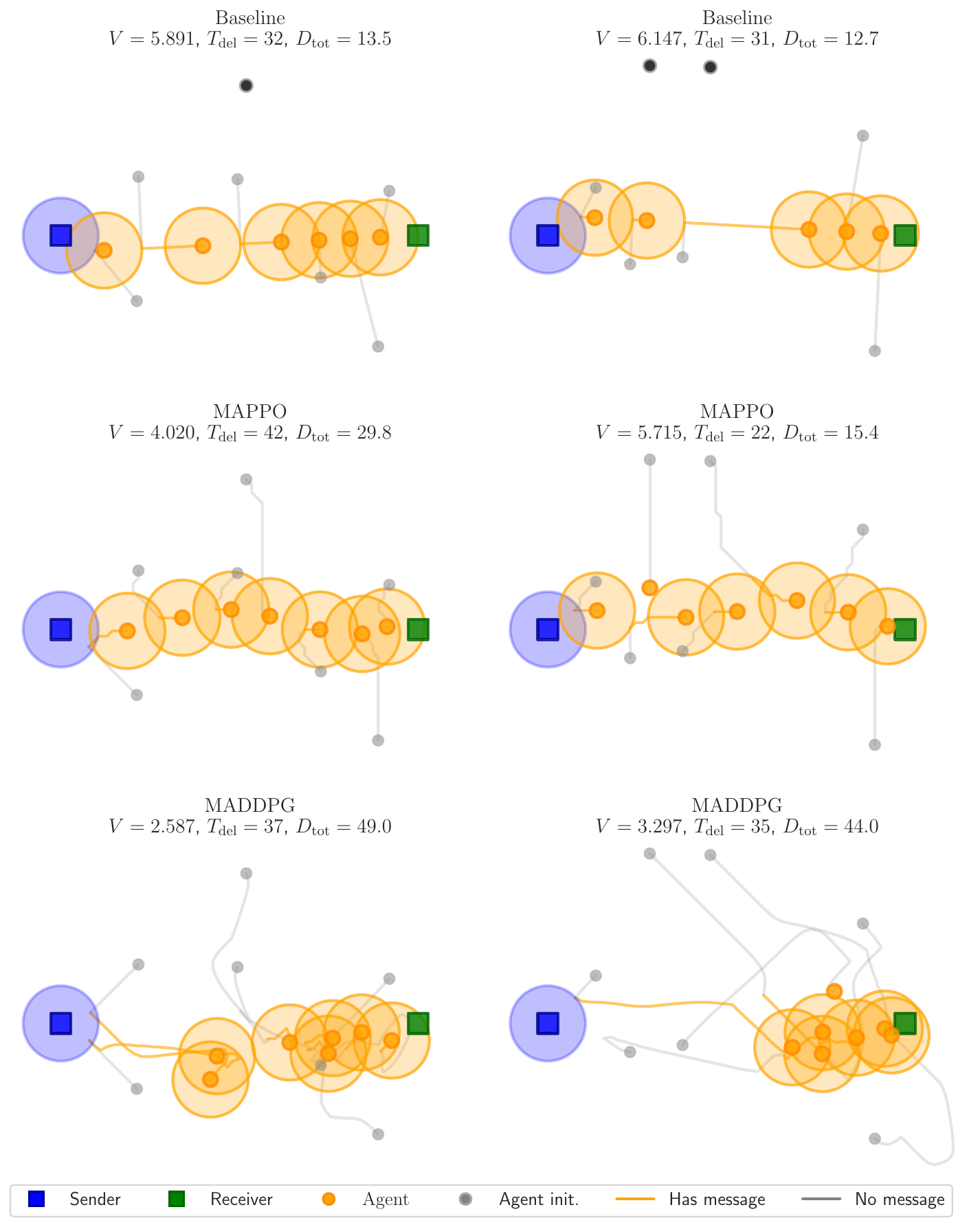}
    \caption{Rollout trajectories from the baseline (top), MAPPO (middle), and MADDPG (bottom) for the the isotropic non-jammed scenario with $K=7$ agents and two initial states, one per column.}
    \label{fig:trajectories_1}
\end{figure}

\begin{figure}[t!]
    \centering
    \includegraphics[width=\columnwidth,
    trim={0cm 0.25cm 0cm 0.27cm},
    clip]{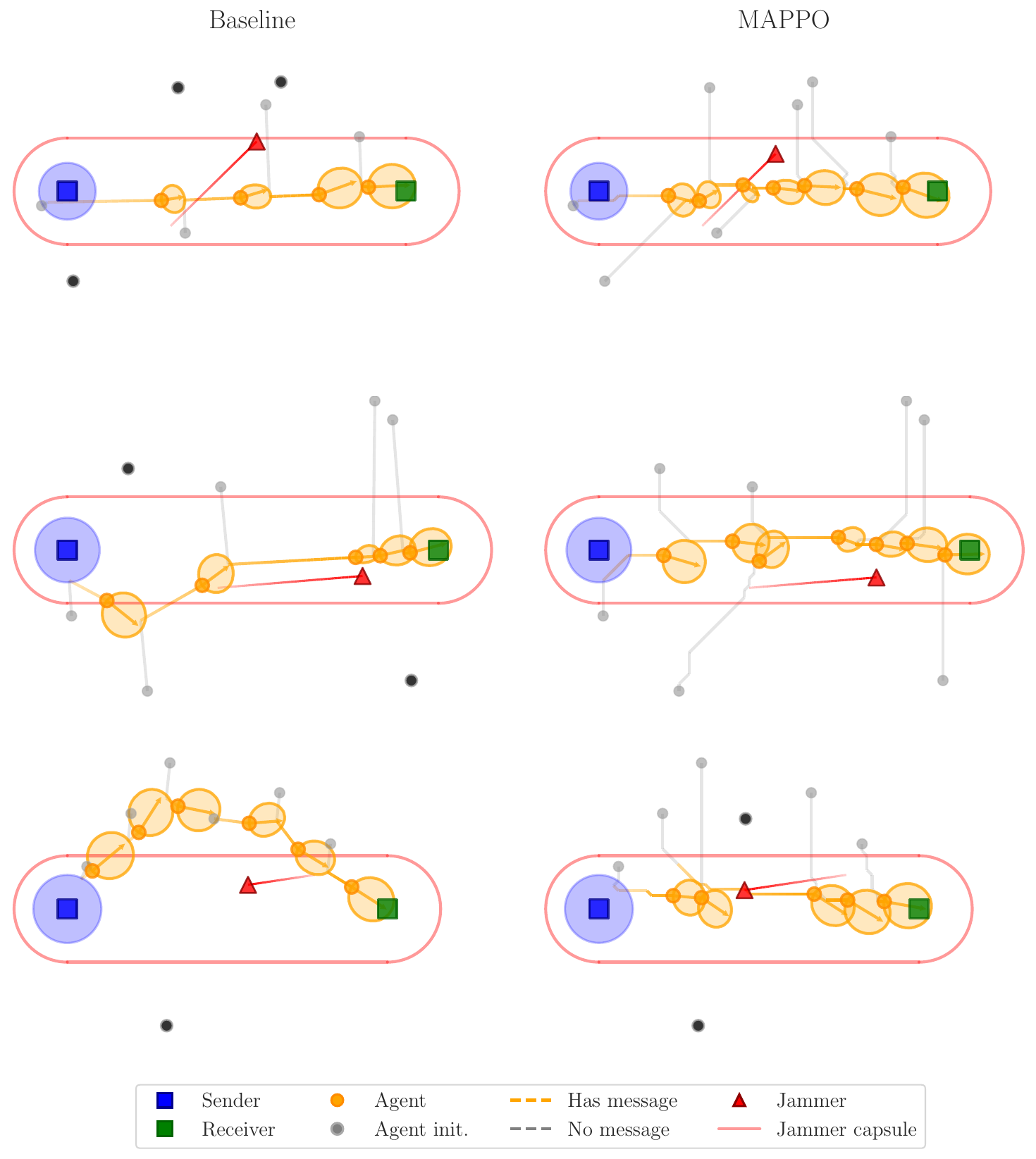}
    \caption{Rollout trajectories from the baseline (left) and MAPPO (right) for the the directional transmission and jammed scenario with $K=7$ agents and three initial states, one per row.}
    \label{fig:trajectories_2}
\end{figure}

Finally, to illustrate the qualitative behavior of the agents under the three policies, Fig.~\ref{fig:trajectories_1} presents rollout trajectories for the isotropic non-jammed scenario with two distinct initial states and $K=7$ agents. MAPPO yields less conservative motion than the baseline while avoiding the unmotivated movements seen in MADDPG. Fig.~\ref{fig:trajectories_2} depicts rollout trajectories for the directed and jammed scenario. These results similarly show that MAPPO is less conservative than the baseline in agent motion, although in the bottom-right case MAPPO has two passive agents.

\section{Conclusion and outlook}
\label{sec:conclusions}

A model problem for collaborative deterministic swarm games with perfect information in discrete time was introduced. By scaling the rewards of the game to allow fair comparisons as the number of agents increases, this model family provides a suitable basis for future scaling studies in Multi-Agent Reinforcement Learning (MARL). 
A robust baseline policy was also proposed, enabling evaluations not only between trained policies but also to a reliable reference.  
Experimental results indicate that Multi-Agent Proximal Policy Optimization (MAPPO) can be trained successfully for a small number of agents, but that scalability issues arise already at nine agents. Multi-Agent Deep Deterministic Policy Gradient (MADDPG) was trained on the easiest of the four scenarios but proved more difficult to train and gave inferior results. 
However, this limited study does not rule out the potential for MADDPG to perform better, and further investigation is required.
Related work has introduced modifications of MAPPO and MADDPG that improved performance \cite{Zhang2020, Bai2024, Zhu2021}, suggesting that tailoring MARL algorithms more closely to the specified problem could be beneficial.
A critical step toward a realistic setting is the introduction of multiple messages and partial observability by letting agents sense the environment. This includes sensing unknown jammers. However, this poses a significant challenge from both a model and algorithm perspective. Additional realism can be achieved by introducing environmental obstacles and implementing collision avoidance.

\bibliography{ifacconf}

\end{document}